\documentclass[twocolumn,amssymb,aps,graphicx]{revtex4}
\usepackage{graphicx}
\usepackage{epsfig} 
\usepackage{amsmath}
\begin{document}
\title{    Delocalization of  Disturbances and the Stability of 
 AC Electricity Grids  }
\author{{\large
 Stefan Kettemann}
}
\address{ Jacobs University,  Department of Physics and Earth Science,
  Campus Ring 1, 28759 Bremen, Germany  }
  \address{
   Division of Advanced
  Materials Science Pohang University of Science and Technology
  (POSTECH) San 31, Hyoja-dong, Nam-gu, Pohang 790-784, South Korea }
        
\begin{abstract} 
 In order to  study  how 
   local disturbances 
affect  the  AC grid stability,
  we start from  nonlinear 
 power balance equations and  map them  to  complex linear wave equations.
 Having obtained  stationary solutions  with phases $\varphi_i$
   at  generator and consumer nodes $i$, we   next  study   the dynamics of  deviations.
         Starting with an initially   localized perturbation, 
         we  find it to spread in a periodic grid 
           diffusively throughout the grid and give
           the parametric dependence of  diffusion constant $D$. 
             We   apply the same  solution  strategy  to general grid topologies and 
             analyse their  stability against local perturbations. 
              The perturbation remains 
               either  localized or becomes delocalized, 
                 depending on  grid  topology, power capacity and distribution of consumers and generator power $P_i$. 
               Delocalization  is found to 
                increase the lifetime of  perturbations and thereby their 
                  influence  on  grid stability, while localization results in an exponentially
                   fast decay of  perturbations at all grid sites.
        These results  may therefore lead to  new strategies to 
        control the stability of electricity grids. 
\end{abstract}

\date{\today }
\maketitle

  The stability of electricity grids
  requires to protect them against   perturbations\cite{kundur2,kundur,machowski}.
     Therefore,  electrical power systems must be 
      constructed in such a way that   a 
       physical disturbance   does not result in 
        exceeding bounds of system variable fluctuations. 
          The  energy transition  towards an increased 
          supply of  decentralized
             renewable energy  necessitates to study 
              consequences of such  structural changes for the 
               stability of electricity grids and  to find efficient 
                ways to modify  them to ensure their stability \cite{amin}.
Since this is a highly complex and nonlinear problem
  the study of its  dependence on  network topology,  operating conditions and 
    forms of disturbances requires to make modeling assumptions\cite{kundur}. 
 Recently, 
 the  synchronization of  rotor angles in   electricity grids has  been  modeled 
 by a network of  nonlinear oscillators\cite{hill,timme,schmietendorf}.
     Here, networks of generators and engines 
     are described by a system of      coupled  differential  equations
        for   local rotor angles of generators and loads $\varphi_i$,
         where  $i$  are grid nodes. 
      The numerical  solution of these differential equations
       showed that, on the one hand,  networks
           become more unstable with increasing decentralization  
            against perturbations on short time 
             scales with large amplitude, while 
     the danger of a blackout 
     can be reduced  by decentralization  \cite{timme}.
  In this article, we 
 study how phase perturbations evolve with time in  AC grids.
 The origin of such  phase perturbations may arise for example due to  local fluctuations in 
   generating power 
   from a wind generator. 
 We start by  finding  stationary solutions for the spatial distribution 
   of  phase $\varphi_i$,  for given distribution of 
  active and reactive power, $P_i$ and $Q_i$ at the
   grid  nodes $i$. Next, we 
   reconsider the nonlinear dynamic power balance equations. 
    For small  deviations from the stationary solutions, we 
   derive  linear wave equations describing  the   
 phase perturbation dynamics.
     Solving these equations, we  explore how  a local  phase perturbation
      propagates with time through the grid. 
        Depending on the geographical  distribution of power,
         grid power capacity  and  topology we find that 
          it may either spread diffusively or    become localized.
        This phenomenon is generally known as  Anderson localization\cite{anderson}, 
     where the coherent scattering of waves in a random medium causes their localization.

{\it  Steady state power flow  in AC transmission grids.---}
       The  power balance equations in AC transmission grids 
       are obtained  from Kirchhoff's laws 
       at node $i$ as
       \begin{equation} \label{steadystate}
        S_i =\sum_j V_i \left(\frac{V_i-V_j}{Z_{ij}} \right)^*,
       \end{equation}
       where $ S_i = P_i + \i Q_i$.  $P_i$ is the active power produced
    at   generator nodes $P_i > 0$, or consumed 
    at consumer nodes $P_i < 0$, satisfying  total power balance
     condition $ \sum_{i=1}^N P_i =0$ with $N$  the total number of nodes.
     $V_i$ is the voltage at node $i$. 
        In an AC grid the reactive power $Q_i$  of  consumers is
                given
               while the one of  generators is adjusted\cite{machowski,heuck}.     
  The   transmission line from node $i$ to $j$  has  impedance $Z_{ij}$. 
         Neglecting  small  losses due to Joule's heat, 
                  we assume them to be purely inductive $Z_{ij} = \i \omega L_{ij}$,
           where  $L_{ij}$ is the transmission 
            line inductance between nodes $i$ and $j,$   $\omega$  the grid frequency. 
             Then,   the voltage at node $i$ is 
                $V_i = V \exp(\i \varphi_i )$, where $V$ is  fixed to  nominal grid voltage and 
           the  power capacity  of a transmission line  is 
               $K_{ij}= V^2/(\omega L_{ij}) A_{ij}$, where $A_{ij}$ is
                the adjacency matrix of the grid.  Note that 
                only  $N-1$ phase angles  $\varphi_i$ remain to be determined
                 as function of the distribution of  power $P_i$ at  $N$ nodes 
                 constrained by the total  power balance condition, so that  
                  reactive power $Q_i$ is fixed  at all nodes. 
               Defining $\psi^0_i (t) =  \exp(-\i \varphi^0_i(t))$, 
               with  $\varphi^0_i(t) = \omega t + \theta^0_i$, 
               phase angle  $\theta^0_i$  at node $i$ in steady state,
               it is  convenient 
               to  write Eq. (\ref{steadystate})
                as a linear wave equation,
                 \begin{eqnarray} \label{stst2}
           S_i  \psi^0_i(t) 
               = \sum_j \i K_{ij} (\psi^0_i(t) - \psi^0_j(t)). 
               \end{eqnarray}
                
{\it Phase dynamics.---}
   Phase  dynamics in AC electricity grids  has been modeled by active power balance equations with  additional  terms  describing the dynamics of rotating 
    machines\cite{hill,heuck,filatrella,timme,schmietendorf,rohden,menck,rohdenchaos,deb}.
   One term describes the inertia to changes of 
     kinetic energy of 
     a synchronous rotating generator or motor with rotor angle $\varphi_i$ 
  with    inertia $J$, when 
     we assume that all loads are either synchronous or induction motors, whose dynamics
      can be modeled that way\cite{machowski}. 
      Another term  describes the damping
     with  coefficient $\gamma$.  
     Adding these  terms to the active power balance equations, the real part of Eq. (\ref{steadystate}),
      yields for purely inductive transmission lines, 
     \cite{hill,heuck,filatrella,timme,rohden,menck}
            \begin{equation} \label{dynamic}
        P_i =\left( \frac{J}{2}  \frac{d}{dt}  + \gamma \right)  \left(\frac{d \varphi_i}{d t} \right)^2  + \sum_j K_{ij} \sin (\varphi_i-\varphi_j).
       \end{equation}
We note that  
   for fixed  voltage $V$ there are 
     no dynamic terms  in the  reactive power balance equation,
     which appear only  in higher order 
      when also voltage dynamics in addition to the phase dynamics 
        is considered\cite{reactive}.

{\it Dynamics of Disturbances in the Grid.---}
 In order to study the propagation of disturbances, such as 
  fluctuations in power supply $\delta P(t)$ or in power capacitance
  $\delta K_{ij}$, we set  $ \varphi_i(t) =  \omega   t + \theta^0_i +  \alpha_i(t)$ 
 with steady state phases  $\theta^0_i$, the
solutions of Eq. (\ref{stst2}). We 
   study the dynamics of   phase disturbances   $\alpha_i(t)$ which are 
                governed by 
                   \begin{eqnarray} \label{alphaharmonic3}
             \partial_t^2   \alpha_i  + 
                  2 \Gamma  \partial_t   \alpha_i   = 
        \frac{P_{i}}{J \omega} -     \sum_j   \frac{K_{ij}}{J \omega} \sin   (\theta^0_i
       - \theta^0_j + \alpha_i -\alpha_j),
          \end{eqnarray}            
          where $\Gamma = \gamma/J$.
          Considering  small  perturbations from  the  stationary state  $\psi^0_i$, 
           we  expand  Eq. (\ref{alphaharmonic3}) in $\alpha_i -\alpha_j$,
        yielding  linear wave equations on the grid, 
                      \begin{eqnarray} \label{alphaharmonic4}
                   \partial_t^2   \alpha_i  + 
                   2 \Gamma  \partial_t   \alpha_i 
                   = -
       \sum_j   t_{ij}    (\alpha_i -\alpha_j),   
          \end{eqnarray}
         with  hopping amplitude $t_{ij} = K_{ij} \cos(\theta^0_i - \theta^0_j)/(J \omega)$.
         Note that $t_{ij}$ depends both on power capacitance $K_{ij}$ and 
         thereby the  grid topology, as well as on 
            the initial distribution of power $P_i$ through the stationary phases $\theta^0_i$.
          With  $\alpha_i (t)   =  \sum_n 
               c_{ni} \exp (-\i \omega_n t)$
                we get   for $\Gamma =0$ the spectral representation of the linear wave equation
                \begin{eqnarray} \label{randommatrix}
                    \omega_n^2   c_{ni}  
                   =      
       \sum_j   t_{ij}    (c_{ni} -c_{nj}),
          \end{eqnarray}               
        with    eigenfrequencies $\omega_n$ and
               eigenmodes $c_n$. $c_0=0,$ $\omega_0 =0$  correspond to the stationary solution. 
               For $\Gamma \neq 0$ we get from
                Eq. (\ref{alphaharmonic4})
               the same eigenmodes $c_n$ with
                complex eigenfrequencies $\Omega_n =  -\i \Gamma
                + \i \sqrt{\Gamma^2 - \omega_n^2 }$.
                  Eq. (\ref{randommatrix}) can be solved for 
                  arbitrary electricity grids like the 
                   german transmission grid shown in Fig. 1 (right), 
                   where the hopping matrix elements $t_{ij}$
          with stationary phases $\theta^0_i,$ are obtained from the solution of 
                   Eq. (\ref{stst2}).

\begin{figure}[Ht!]
    {\includegraphics[width=0.23\textwidth]{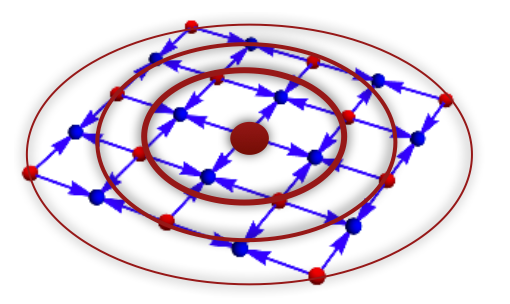}}
    \includegraphics[width=0.23\textwidth]{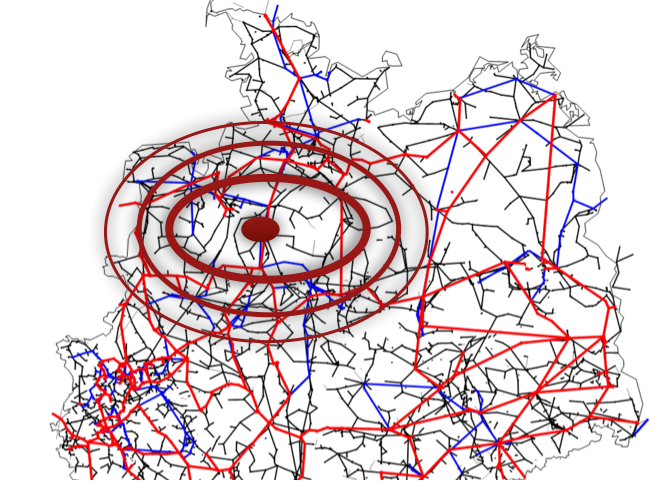}
  \caption{ Left: Square grid of  generators (red) and consumers (blue). Arrows: direction of  active power transmission $F$.     Right: Topology of  german transmission (380kV(red), 220kV(blue)) and high voltage distribution grid (110kV(black))\cite{samyak}.   A propagating 
  disturbance is sketched as red circles.    }    
  \label{fig:quasi_dos}
\end{figure}

              {\it  Square grid.---} 
            As an example,  let us first consider 
                 a grid where all transmission lines have 
          equal length $a$.  We start with 
                 a periodic arrangement 
                of generators and consumers as shown for a square
                grid in Fig. \ref{fig:quasi_dos} (left). 
      Assuming that all generators on  sublattice G
      generate  power  ${P}_{{\bf x}} = +P$,  ${\bf x} \in G$, while 
        all consumers on sublattice C  consume  power ${P}_{{\bf x}} = -P$,  ${\bf x} \in C$,
         we find the solution by making the 
         Bloch Ansatz  for the stationary solution 
          \begin{eqnarray} \label{static1}
          \psi_k ({\bf x}\in G,t) &=& \psi_{G k} e^{\i{\bf k x}} \exp (-\i \omega t),
          \nonumber \\
            \psi_k ({\bf x}\in C,t) &=& \psi_{C k} e^{\i{\bf k x}} \exp (-\i \omega t),
          \end{eqnarray}
           where ${\bf k}$ is a wave number. 
          For periodic boundary conditions
            on the grid of linear size $L$ in all $d$ directions  with   
             unit vectors $\hat{e}_n$, $n=1,...,d$, $\psi_k ({\bf x}+ L \hat{e}_n) =\psi_k ({\bf x})$,
              for all $n=1,...,d,$ the wavenumber is   ${\bf k} = 2 \pi {\bf n}/L,$
               where the components of the vector $\bf{n}$, $n_{n}, n=1,...,d$  are integers.
      For each wave number ${\bf k}$ we find a solution of the form Eq. \ref{static1}, where 
    the phase factors  
         $ \psi_{G k},    \psi_{C k}$
          are  given   by            \begin{equation} \label{static2}
             \psi_{C k} = \exp \left( \i  \delta_k \right) \psi_{G k} {\rm ~with~}  \sin  \delta_{k} =   P/(f_k K),
             \end{equation}  
                   where  $f_k = 2 \sum_{n=1}^d \cos (k_n a)$   
                    depends on wave number components  $k_n$. 
           For  given  reactive power $Q$, which  is for this arrangement of consumers
            and generators constrained by 
             Eq. (\ref{stst2}) to be  the same
             at all nodes, the
           wave vector ${\bf k}$ is   determined by the equation
                 \begin{equation} \label{static3}
                f_k^2   = (Q/K - 2 d)^2 + P^2/K^2.
                  \end{equation}

%

             The transmitted power
                  between sites $i$ and $j$ is  $ F_{ij} = \i K_{ij} (\psi_i(t) - \psi_j(t)) \psi_i(t)^*$.
                  For the homogenous state ${\bf k}=0$ 
                    the  active power transmitted between 
                           neighbored  sites $i \in G$ and  sites $j \in C$ is 
                              $Re F_{ij} =  P/(2d),$ as shown 
 in  Fig.  \ref{fig:quasi_dos} (left)  on a square lattice, where arrows indicate 
  the direction of  active power transmitted from  generators  to
   neighbored consumers. The  reactive power in the transmission lines   is
   $Im F = K (1-\sqrt{1-\frac{P^2}{4 d^2 K^2}})$, which is for $ P\ll K$
   much smaller than the transmitted active power.
    For finite  wave vector ${\bf k}$ 
     the   active power transmitted  between  $i \in G$ and $j \in C$ 
       for ${\bf k} = k  \hat{e}_n$ is 
      $ Re  F_{ij} = K \sin (\delta_k \pm k_n a)$,
      with unit vector between $i$ and $j,$  $\hat{e}_{ij}  = \pm    \hat{e}_n.$
  In all other directions $ Re  F_{ij} = K \sin \delta_k$. 

  {\it Stability.---}
Disregarding the dependence of  
 $\alpha_i$ on the perturbation at neighbored sites $\alpha_j$ 
reduces   Eq. (\ref{alphaharmonic3})
 to the  one of  a damped, driven nonlinear pendulum.
                For large times $t \gg 0$ it has   two stable solutions:               
                 {\it 1.}  {\it 
                  Stationary solution} $\partial_t \alpha_i =0$, 
                   $\alpha_i = n 2 \pi,$ $n$ integer,
                    to which  small 
        deviations decay exponentially.                      
{\it 2.}   {\it  Overswinging pendulum solution,} when 
  driving force and   damping are
 in balance.
The phase velocity   converges then to
                       \begin{eqnarray} \label{converge}
                      \partial_t \alpha_i(t)       = \Omega_i 
                          -\sum_j  \frac{K_{ij} \sin (\Omega_i t +\theta_i^0 -\theta_j^0 +\eta_i)}{J \omega  \sqrt{\Omega_i^2 + 4\Gamma^2 }},          \end{eqnarray}  
        with frequency  $\Omega_i = P_i/(2 \Gamma  J \omega)$ and phase shift $\eta_i= \arctan (2 \Gamma/\Omega_i)$.   
        In Fig.  2    phase curves are  shown for the square lattice
                with  $\theta_i^0 -\theta_j^0  =\pm \delta_k +k_n a$. Regions of  
         stability   are  shaded in blue, outside which
         all  phase points 
         converge  to  
   open orbit solution    Eq.  (\ref{converge}) (red).  
    A  condition   for phase points to lie inside the stability region  is obtained 
    by  approximating its irregular shape   by an ellipse whose vertex  on the $\alpha$-axis is given by the saddle point,
     $\alpha_{i l } = 2 \delta_k-\pi$. Its
 vertex on the $\partial_t \alpha$-axis is given 
   by a  linear interpolation of  stable trajectory on the separatrix 
   $\partial_t \alpha = -   (\pi - 2 \delta_k ) \tan \zeta$,  where 
 $\tan \zeta =   -\Gamma/ \Omega_i  -
         \sqrt{\Gamma^2/ \Omega_i^2  +  \frac{d K \cos (  2 \delta_k)}{J \omega \Omega_i \sqrt{\Omega_i^2 + 4\Gamma^2 }} }. $   Thus,  phase points are  inside the basin  of attraction if they 
      satisfy
      \begin{equation} \label{basinofattraction}
   \alpha_i^2 + \frac{(\partial_t \alpha_i)^2}{\tan^2 \zeta_i} \ll (\pi - 2 \delta_k)^2.
    \end{equation}
 
\begin{figure}[Ht!]
  {\includegraphics[width=0.2\textwidth]{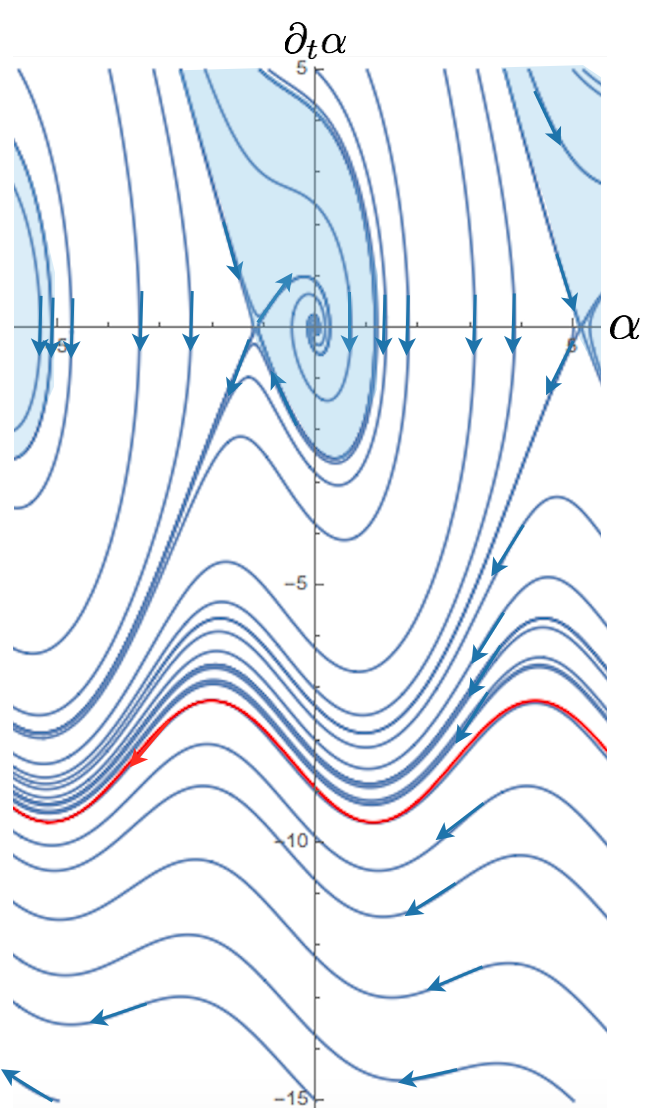}}
  \caption{  Phase curves of Eq. (\ref{alphaharmonic3}) for $\alpha_i$,
   setting $\alpha_j=0$.
    Flow directions  are indicated by arrows.
     Stable fixed points are $\alpha_i = n 2 \pi, \partial_t \alpha_i =0.$  
       Separatrices are $ \alpha_{i l} =2  \delta_k +  (2 l+1) \pi,  \partial_t \alpha_i =0.$ 
       Blue shaded areas are basins of attraction. All other phase points converge  to  
   Eq. (\ref{converge}) (red curve). 
       }    
  \label{fig:phasediagram}
\end{figure}
      
        
                {\it Propagation of Local Disturbances.---}
             Now, we can study the propagation of disturbances in a square grid by inserting 
     the analytical  steady state  solutions Eqs. (\ref{static1},\ref{static2},\ref{static3})
     into the dynamical equations Eq. (\ref{alphaharmonic3}).             
                    If we   perturb the phase 
                       at node ${i=n}$, 
                     $\alpha_n(t=0) = \alpha_0 \neq 0$, 
              while    $\alpha_i(0) = 0$ 
              at all other nodes $i \neq n,$
              that disturbance    excites nodes
$i \neq n$ at later times $t>0.$
 For a small perturbation, satisfying  Eq. (\ref{basinofattraction}), 
     we  can use  linearized wave equations Eq. (\ref{alphaharmonic4}) 
     with $t_{ij} =
   K_{ij} \cos \delta_k/(J \omega)$. 
          Using the spectral representation
          $\alpha_i(t) = \sum_q c_q e^{\i {\bf q r}_i}   e^{-\i \epsilon_q t},$
    insertion into Eq. (\ref{alphaharmonic4}) gives the complex frequency 
      \begin{eqnarray} \label{alphaharmonic2}
            \epsilon_q
                   =- \i \Gamma \left( 1 \pm \sqrt{1 - \frac{2 d K \cos \delta_k}{J \omega \Gamma^2 } (1 -\frac{f_q}{2d})} \right).
          \end{eqnarray}
       For finite $\Gamma$ and  small $q$ 
    the dispersion is quadratic,  $  \epsilon_q
                   =- \i \frac{ K  a^2 \cos \delta_k }{J \omega \Gamma} {\bf q}^2.$   
                   For large  
                  momenta $q$, the ballistic limit, 
           the dispersion is linear
            $   \epsilon_q|_{\Gamma=0} = v_k q$ with velocity $v_k = \sqrt{K \cos \delta_k/(J \omega)} a$. 
    An initially localized perturbation,  $ c_q =const.$
becomes
for  times $ t> \tau = 1/\Gamma$ and distances exceeding the mean 
 free path  $l = v \tau$, $|r_{\rm i} -r_{\rm l}| > l$, 
       \begin{eqnarray} \label{diffusion}
             \alpha_i (t)
                   =\frac{ \alpha_0 }{(4 \pi D_k t/a)^{d/2}} \exp \left( - \frac{({\bf r_{\rm i} -r_{\rm l}})^2}{4 D_k t} \right).
          \end{eqnarray} 
Thus, the initially localized perturbation  spreads diffusively with  diffusion constant 
      \begin{equation} \label{diffconst}
       D_k =  \frac{ K a^2 \cos \delta_k  }{\omega \gamma } = \frac{v_k^2 \tau }{2},
       \end{equation}
       as shown in Fig. 3, where 
             $\partial \alpha_i$  is plotted versus
                 phase perturbation   $\alpha_i$ 
                  at initial site $i=l$ and  other sites $i',i''$ for
                   diffusive times $t > \tau = 1/\Gamma$,    using Eq. (\ref{diffusion}). 
                Time is progressing as indicated on the color bar in units of
    $\tau$. 
   Diffusion   causes slow decay of the perturbation 
      at the initial site, and  an initial  increase,  followed by a slow decay at  other sites.
      The area $A_t$   where  nodes become 
       perturbed at time $t$, increases diffusively  with time as $A_t = D_k t$. 
  \begin{figure}[t]
  {\includegraphics[width=0.45\textwidth]{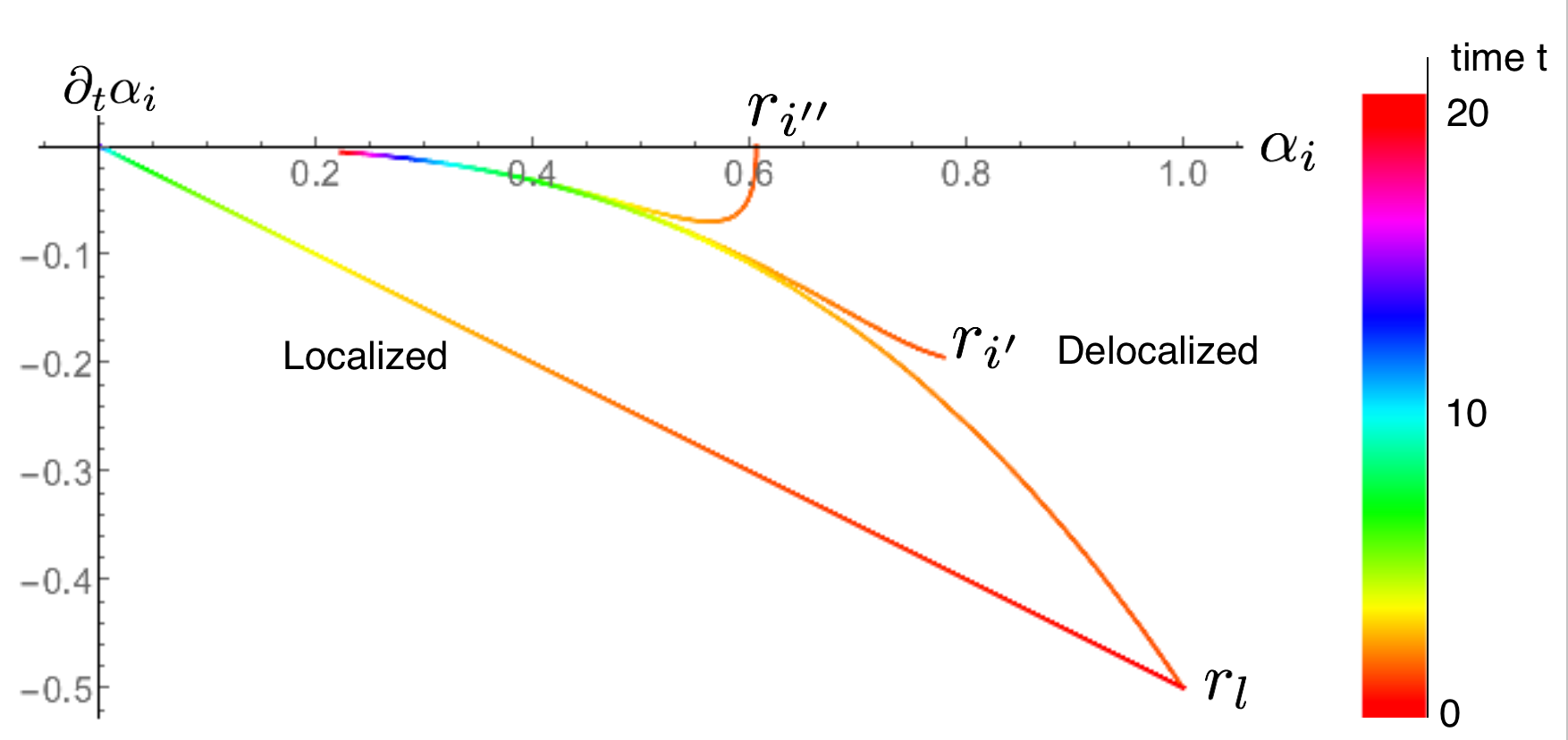}}
  \caption{(color online)   
  Phase curves  at initial site $r_l$ and  other sites $r_{i'},r_{i''}$. Time  progresses as indicated in  color bar in units of
    $\tau = 1/\Gamma$.
   Delocalization  causes  slow decay of the perturbation 
      at the initial site, and  an initial  increase,  followed by a slow decay at   other sites (upper curves).
  Localization causes an exponentially  fast decay  at all sites  (lower curve).  }    
  \label{fig:deloc}
\end{figure}
    In order to quantify  the stability with 
      Eq. (\ref{basinofattraction})
            we calculate  the spatial average,
     $     \delta \alpha = \sqrt{ \frac{1}{N} \sum_i \alpha_i^2}=
          \frac{\alpha_0}{ (8 \pi D_k t/a)^{d/4}}, 
                \delta \omega = \sqrt{ \frac{1}{N} \sum_i  (\frac{d\alpha_i}{dt})^2}= \frac{\sqrt{d(2+d)} }{4  t} \delta \alpha$,
  decaying with a power law in time.
                For 
                  power capacity  $K = .5 GW$, $d=2$, grid  frequency $ \omega = 2 \pi ~ 50/s$, 
           power $ P = 1.9 GW$, inertia  $J = 10^5 kg  m^2$
           and damping  $\Gamma = 1/s$  the
            perturbation spreads initially  with  velocity $v=2.22 a/s$. 
            Beyond 
                mean free path $l =2.22 a$ it spreads diffusively with 
                 diffusion constant     $D_0 =  2.46 a^2 /s$. 
                  The time to diffuse a 
                     length  $L=10 a$ is  $t_D = L^2/D_0 =  40.64 s.$
 The closer power $P$ is to capacity limit $K_c = 2 d K$,  the smaller is  diffusion constant $D,$
   the longer survives the disturbance.  
  
         The propagation of 
         any type of disturbance can be studied with Eq.  (\ref{alphaharmonic3}).
          F. e. a change in power $\delta P$  at  $t_0=0$
           at  neighbored nodes $i,j,$ $\delta P_i = - \delta P_j  = \delta P$
           changes  transmitted power between nodes $k,l$ at time $t$,
           \begin{equation} \label{changeFdynamic}
           \delta F_{kl}(t) = \pm  \delta P A_{kl} \frac{\pi^2 a^2  }{\omega_0 D_k t^2} \exp (-\frac{({\bf r_{\rm i} -r_{\rm l}})^2}{4 D_k t}),
           \end{equation}
           with $+$ when $k$ is a generator and $l$ a consumer and $-$ vice versa.
      Similarly,    the change in power flow due to a 
       static perturbation $\delta P_i = - \delta P_j  = \delta P$ is  given by 
        \begin{equation} \label{changeFstatic}
           \delta F_{kl} = \pm \delta P A_{kl} 2 \pi  a^2 /({\bf r_{\rm i} -r_{\rm l}})^2.            \end{equation}
           An instantaneous change in power capacitance $ \delta K A_{ij}$ causes a change of power
           Eq. (\ref{changeFdynamic}),
             and 
             Eq. (\ref{changeFstatic}) for a static perturbation,  replacing
              $\delta P$ by $\delta K,$ respectively.            
      
          {\it Localization  of Disturbances.---}
       Eq. (\ref{randommatrix}) can be applied to 
        study the propagation of disturbances in  AC  grids of 
        arbitrary  distribution of 
     power $P_i,$  arbitrary grid  topology and power capacity $K_{ij}$ 
      by calculating  the 
         hopping amplitude $t_{ij} = K_{ij} \cos(\theta^0_i - \theta^0_j)/(J \omega)$
          from the steady state solutions $\theta^0_i$, as found by solving Eq.  (\ref{dynamic}). 
        In a first attempt to model   the complexity of a real grid, 
       we   take  a random distribution of $t_{ij}$
        caused by  the  wide distribution of 
     power $P_i$ and the real grid  topology 
     with its complex network structure.
     Eq. (\ref{randommatrix}) first   appeared  in the problem of randomly coupled atoms
                 in  harmonic approximation.     
          For chains it has been solved
                  for various random distributions of $t_{ij}$\cite{dyson,alexander,ziman,wegner}.            
          If  $t_{ij}$ is  taken   from a box distribution, 
              the density of eigenmodes is  
                 constant, $\rho(\Omega_n)= 1/\omega,$
                  for $0 < \Omega_n < \omega$. For nonzero $\Omega_n$ 
               the eigenstates are   localized with  localization length  
                $ \xi(\Omega_n) \sim 1/\Omega_n$ \cite{dyson,alexander,ziman,wegner}. 
            The localization length  of the lowest  eigenfrequency  
                $\Omega_1\sim \omega  a/L$   is large, of the order
                 of the system size $\xi_1 \sim L$. 
                   The highest eigenfrequency 
                   $\Omega_n \rightarrow \omega$ has 
                    a  localization length $\xi \rightarrow 2 a$, twice the length  $a$ 
                  of a transmission line.    In dimension $d=2,$ which is the situation most 
                  relevant for real grids, 
                   all eigenstates remain localized for nonzero $\Omega_n$, but the 
                    localization length can be for small $\Omega_n$ exponentially large, 
                     so that it typically is larger than the system size for realistic grid extensions $L$. 
                      This might explain that in Ref. \cite{jung} we found in square 
                       grids  a 
                      long range power law decay with power $q=2$  as in Eq. 
                      (\ref{changeFstatic}), even when taking a random distribution of $P_i$. Also in a real grid topology we found long range decay\cite{jung}.                     
                      Typically, the localization length is smallest in treelike grids, while it becomes 
                       larger, the more  meshed the grid becomes. 
                   In  dimensions $d > 2$, 
                   there is a critical value $\omega_c$, such that for
                    $\omega_n > \omega_c$ all modes are localized, while they are extended for
                      $\omega_n < \omega_c$ \cite{john,wegner}.
                       If the phase perturbation  is initially in  a  state localized 
                       around site $r_0$ 
                        with localization length $\xi_n$, it 
                        decays in time $t$ as $\alpha_i(t) = \alpha_0 \exp(- \frac{\mid r_i -r_0 \mid}{\xi_n})
                         \exp(- \tilde{\Gamma}_n t),$
                 where $\tilde{\Gamma}_n  =Re [ \Gamma -  \sqrt{\Gamma^2 - \omega_n^2 }]$.
                The typical phase and frequency shift is  found to 
                decay exponentially fast as 
                \begin{equation} \label{deltaloc}
                 \delta \alpha  = \alpha_0 \sqrt{\xi_n/a}
                         \exp(- \tilde{\Gamma}_n  t), \hspace{.5cm}
                   \delta \omega  =   \tilde{\Gamma}_n  \delta \alpha.
                \end{equation}
              Thus, localization  causes  exponentially fast 
               decay of phase perturbations at all nodes $i$,
               as shown in Fig. 3, where it is  compared with a delocalized phase perturbation
                decaying with a power in time $t$, Eq. (\ref{diffusion}).

              
         {\it Conclusions.---} 
          Local  perturbations, 
      arising for example from power fluctuations, 
           are found to spread diffusively in a periodic grid,
           decaying slowly with a power law in time and space.
       The closer the generator power $P$ 
       comes  to the capacity limit $K_c$,  
        the smaller is  diffusion constant $D$ and the  longer 
         takes  the  perturbation to decay.  
            Modeling the complexity of  a realistic grid with a
           random distribution of  generators and consumers and/or random transmission power capacity
             the  phase perturbation is found to   become  localized in 1- and 2-dimensional  grids.  
          Localization   leads to an exponentially fast decay of  phase perturbations at all sites, while delocalization results in diffusive slow decay, 
               Fig. \ref{fig:deloc}.
                 Initially 
                      small  perturbations  may  then add up 
                       at some nodes to large perturbations and  push the system 
                        outside  of the region of stability. We  conclude that  it  is favorable 
                  for   stable grid operation to ensure that 
             phase perturbations remain localized, decaying
                     exponentially fast  at all sites.        
                                     The consequences of these results for real electricity grid topologies will 
                        be studied by solving Eqs. (\ref{stst2},\ref{randommatrix}) numerically
                         in a future publication.  We  also plan to study how the spreading of perturbations is modified when including voltage fluctuations, using for example the third order model\cite{kundur2,machowski,schmietendorf}.  
                         While we  assumed  here a network of synchronous generators and motors,                                  modern wind turbines  are rather  induction generators, the most modern ones being the doubly fed induction generator, converting the power from 
 AC to DC and then to AC with the grid frequency\cite{machowski}.
   Thus,   the  energy transition  towards an increased 
          supply of  decentralized
             renewable energy  necessitates  to get a better understanding 
how  the dynamic equations are modified  and to  understand the resulting  consequences for  the grid dynamics.
                      
%
%
%
%
%
%

\acknowledgments
We thank M. Rohden  for stimulating discussions and D. Jung for useful comments.
  We acknowledge gratefully  support of BMBF
   CoNDyNet FK. 03SF0472A.


\begin{references}
\bibitem{reactive}
 Reactive power balance dynamics is directly coupled to 
     voltage dynamics
 \cite{kundur2,machowski,schmietendorf}.
\end{references}
\end{document}